\documentclass{PoS}
\usepackage{graphicx}
\usepackage{paralist}
\usepackage{fancyvrb}

\PoS{PoS(LAT2005)104}

\FullConference{XXIIIrd International Symposium on Lattice Field Theory\\
                 25-30 July 2005\\
                 Trinity College, Dublin, Ireland}

\title{Lattice QFT with FermiQCD}

\ShortTitle{Lattice QFT with FermiQCD}

\author{M.~Di~Pierro \\
    School of Computer Science, Telecommunications and Information Systems \\
        DePaul University, Chicago, IL 60604, USA}

\author{J.~M.~Flynn\\
        School of Physics and Astronomy, \\
	University of Southampton, Southampton, SO17 1BJ, UK}

\abstract{FermiQCD is a C++ library for fast development of parallel
  Lattice Quantum Field Theory computations. It has been developed
  following a top-down fully Object Oriented design approach with focus
  on simplicity of use. FermiQCD includes: a heatbath algorithm for
  Wilson and $O(a^2)$ improved $SU(n)$ gauge actions; inversion
  algorithms for Wilson, Clover, Kogut-Susskind, Asqtad, and Domain
  Wall fermionic actions; example programs for various types of meson
  propagators; and converters for the most common gauge file formats.}

\begin{document}

\section{INTRODUCTION}

FermiQCD~\cite{DiPierro1,DiPierro2,DiPierro3} is a library for fast
development of parallel applications for Lattice Quantum Field
Theories and Lattice Quantum Chromodynamics~\cite{DiPierro4}. It was
designed both to be easy to use, with a syntax very similar to common
mathematical notation, and, at the same time, optimized for PC
clusters.

FermiQCD takes a top-down approach: the top level functions were
designed first, followed by optimized implementations of those
functions. The most critical parts are optimized in assembler using
SSE/SSE2 instructions. All FermiQCD algorithms are parallel but
parallelization is hidden from the high level programmer. At the
lowest level, parallelization is implemented in MPI and/or in PSIM.
PSIM is a parallel emulator that allows running, testing and debugging
of parallel code on a single processor PC without requiring MPI.

All components are implemented as separate, although related, classes.
For example, in FermiQCD lattices and fields are objects while actions
and inverters are static methods of the corresponding classes.
FermiQCD components range from low level linear algebra, fitting and
statistical functions (including the Bootstrap and a Bayesian fitter
based on Levenberg-Marquardt minimisation) to high level parallel
algorithms specifically designed for lattice quantum field theories
such as the Wilson~\cite{Wilson} and $O(a^2)$-improved gauge actions,
the Clover fermionic action, the Asqtad~\cite{asqtad} action for KS
fermions, and the Domain Wall action~\cite{kaplan}.

One can create a new action by creating a new class and plugging it
into the rest the library. All other components, such as the
inverters, will work with it. For example FermiQCD provides three
inverters, MinRes, BiCGStab and UML. The first two are general and
work with any action and any type of field, the third (UML) is highly
optimized for KS and ASQTAD actions.

\begin{figure}[b]
\begin{center}
\includegraphics[width=0.5\textwidth]{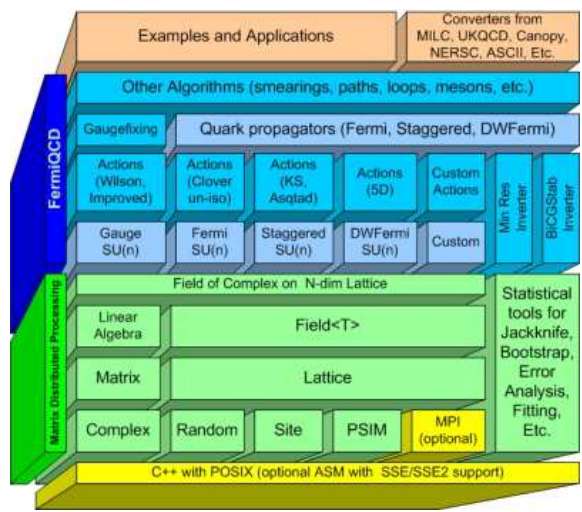}
\end{center}
\caption{Components of FermiQCD.}
\label{fig:cpts}
\end{figure}
Figure~\ref{fig:cpts} shows a schematic representation of FermiQCD's
components. The lower components are referred to as Matrix Distributed
Processing and they define the language used in FermiQCD. The upper
components are the algorithms. The top components represent examples,
applications and other tools. These tools include converters for the
most gauge field formats: MILC, NERSC, UKQCD, CANOPY, and some binary
formats.

\section{SYNTAX OVERVIEW AND PROGRAM EXAMPLE}

All FermiQCD algorithms are implemented on top of an Object Oriented Linear Algebra package with a Maple-like syntax. For example
\begin{eqnarray}
A&=&\mathrm{trace}(\gamma^2 (\gamma^0-\mathbf{1})/2) \\
B&=&e^{i\theta\lambda_3}
\end{eqnarray}
where $\gamma^\mu$ are the Dirac Gamma matrices in Euclidean space and 
$\lambda_3$ is one of the generators of $SU(3)$, are implemented in
FermiQCD as
\begin{Verbatim}[numbersep=3pt]
Complex A = trace(Gamma[i]*(Gamma[0]-1)/2);
Matrix B = exp(I*theta*Lambda[3]);
\end{Verbatim}
A $4D$ $16\times8^3$ lattice is declared (with obvious generalization
to arbitrary dimensions and sizes) as
\begin{Verbatim}[numbersep=3pt]
int L[]={16,8,8,8};
mdp_field lattice(4,L);
\end{Verbatim}
An $SU(n)$ gauge field $U_\mu(x)$ is declared and initialized by 
\begin{Verbatim}[numbersep=3pt]
gauge_field U(lattice,n); set_cold(U);
\end{Verbatim}
The following sets $\beta=6.0$ and performs 10
heatbath~\cite{heatbath} steps with the Wilson gauge action 
\begin{Verbatim}[numbersep=3pt]
coefficients gauge; gauge["beta"]=6.0;
WilsonGaugeAction::heatbath(U,gauge,10);
\end{Verbatim}

Any field can be saved: {\tt U.save("filename");} loaded: {\tt
U.load("filename");} and translated: {\tt U.shift(mu);}. A field can
also be transformed locally. Here is how to implement a global gauge
transformation $G$ of $U$
\begin{Verbatim}[numbersep=3pt]
Matrix G=exp(2*I*Lambda[2]);
mdp_site x(lattice);
forallsites(x)
   for(int mu=0; mu<U.ndim; mu++)
      U(x,mu)=G*U(x,mu)*inv(G);
\end{Verbatim}
{\tt U(x,mu)} is an $n\times n$ matrix and {\tt x} is an object that
represents a lattice site. {\tt forallsites(x)} is a parallel loop.
Each processing node loops over the lattice sites stored by the node.

A Wilson fermionic field is declared as
\begin{Verbatim}[numbersep=3pt]
fermi_field psi(lattice,n);
\end{Verbatim}
and a gauge invariant shift can be implemented as
\begin{Verbatim}[numbersep=3pt]
psi_up(x)=U(x,mu)*psi(x+mu);
psi_dw(x)=hermitian(U(x-mu,mu))*psi(x-mu);
\end{Verbatim}
Notice that {\tt x+mu} reads as $x+\hat\mu$ and {\tt x-mu} reads as $x-\hat\mu$ where $\mu=0,1,2,3$ is one of the possible lattice directions. 

Multiplication by the fermionic matrix is invoked as follows
\begin{Verbatim}[numbersep=3pt]
coefficients quark; quark["kappa"]=0.1245; quark["c_{SW}"]=0.0;
if(quark["c_{SW}"]!=0) compute_em_field(U);
default_fermi_action=FermiCloverActionFast::mul_Q;
mul_Q(psi_out,psi_in,U,quark);
\end{Verbatim}
The chromo-electromagnetic field is required by the clover term and
computed only if required. It is stored inside a gauge field object.
FermiQCD includes three different equivalent implementations of the
above algorithm, declared in the following classes:
FermiCloverActionSlow, FermiCloverActionFast, FermiCloverActionSSE2.
The second is optimized in C++, the third is optimized in assembler.
 
The inverse multiplication
$\psi_\mathrm{out}=Q^{-1}[U]\psi_\mathrm{in}$ is invoked with the
following call
\begin{Verbatim}[numbersep=3pt]
mul_invQ(psi_out,psi_in,U,quark,1e-20,1e-12);
\end{Verbatim}
where {\tt 1e-20} is the target absolute precision for the numerical inversion and {\tt 1e-12} is the target relative precision. 

An ordinary quark propagor is declared and generated by
\begin{Verbatim}[numbersep=3pt]
fermi_propagator S(lattice,n);
generate(S,U,quark,1e-20,1e-12);
\end{Verbatim}
and it can be used, for example, to build a meson propagator 
$C_{\pi}(t)$ by summing the following expression over $x$ and over the
spin components {\tt a}, {\tt b}
\begin{Verbatim}[numbersep=3pt]
Cpi[x(TIME)]+=real(trace(S(x,a,b)*hermitian(S(x,a,b))));
\end{Verbatim}

Everything works similary for the other actions and other types of fields.
Figure~\ref{fig:prog} shows a complete parallel program for generating
{\tt nconfig} gauge configurations in $SU(5)$ on a $16\times 8^4$
lattice, saving them and computing the average plaquette and the pion
propagator on each configuration.
\begin{figure}
\begin{center}
\begin{Verbatim}[numbers=left, numbersep=3pt, xleftmargin=2em,
                 fontsize=\small, baselinestretch=0.97]
#include "fermiqcd.h"
int main(int argc, char **argv) {
  mdp.open_wormholes(argc, argv);   // START
  define_base_matrices("FERMIQCD"); // set Gamma convention
  int n=5, nconfig=100;
  int L[] = {16,8,8,8};
  mdp_lattice lattice(4,L);         // declare lattice
  gauge_field U(lattice, n);        // declare fields
  fermi_propagator S(lattice,n);    // declare propagator
  mdp_site x(lattice);              // declare a site var
  coefficients gauge; gauge["beta"]=6.0; // set parameters
  coefficients quark; quark["kappa"]=0.1234; quark["c_{SW}"]=0.0;
  default_fermi_action=FermiCloverActionFast::mul_Q;
  mdp_array<float,1> Cpi(L[TIME]);    // declare and zero Cpi
  for(int t=0; t<L[TIME]; t++) Cpi(t)=0;
  set_hot(U);
  for(int k=0; k < nconfig; k++) { 
     WilsonGaugeAction::heatbath(U,gauge,10); // do heatbath
     mdp << average_plaquette(U) << endl;     // print plaquette
     U.save(string("gauge")+tostring(k));     // save config
     if(quark["c_{SW}"]!=0) compute_em_field(U);
     generate(S,U,quark,1e-20,1e-12);         // make propagator
     forallsites(x)                           // contract pion
       for(int a=0; a<4; a++)                 // source spin
         for(int b=0; b<4; b++)               // sink spin
           Cpi[x(TIME)]+=real(trace(S(x,a,b)*hermitian(S(x,a,b))));
     mpi.add(Cpi.address(),Cpi.size());       // parallel add
     for(int t=0; t<L[TIME]; t++)    
       mdp << t << " " << Cpi(t) << endl;     // print output
  } 
  mdp.close_wormholes();            // STOP
  return 0;
 } 
\end{Verbatim}
\end{center}
\caption{A complete parallel program in FermiQCD.}
\label{fig:prog}
\end{figure}

\section{BENCHMARKS}

Table~\ref{tab:bmarks} shows typical running times for the FermiQCD
inverters applied to different actions. Times are in microsecond per
lattice site per step. Notice that MinRes involves one {\tt mul\_Q}
per step, BiCGStab involves two, and the UML inverter also involves
two but only applied to sites of even parity. These times were
computed on one 3.2GHz Pentiutm 4 (typical computations in parallel
with a Myrinet network show a drop in efficiency of 20-30\% when
scaling up to 16-32 processors).
\begin{table}
{\footnotesize \begin{center}
\begin{tabular}{|l|l|cccc|} \hline
Action 	& Inverter & float & float (SSE)& double & double (SSE2) \\ \hline
Wilson 	& Min Res & 	8.83 & 	1.79 & 6.84 & 2.07 \\
Wilson 	& BiCGStab & 	17.8 & 	3.16 & 13.8 & 4.42 \\
Clover 	& Min Res & 	9.76 & 	1.98 & 12.08 & 2.82 \\
Clover 	& BiCGStab & 	19.63 & 4.71 & 24.95 & 6.08 \\
KS	& Min Res & 	1.42 & 	0.78 & 1.71 & 1.01 \\
KS	& BiCGStab & 	2.95 & 	1.63 & 3.56 & 2.11 \\
KS	& UML & 	1.89 & 	1.14 & 2.08 & 1.34 \\
Asqtad 	& Min Res &  	3.73 & 	2.47 & 4.29 & 5.24 \\
Asqtad 	& BiCGStab &  	7.65 & 	5.02 & 8.79 & 6.61 \\
Asqtad 	& UML 	&  	1.14 & 	3.14 & 5.24 & 	3.81 \\ \hline
\end{tabular}
\end{center}}
\caption{Running times for FermiQCD inverters using different
  actions.}
\label{tab:bmarks}
\end{table}

\section{CONCLUSIONS}

FermiQCD is now a stable and mature product and the project mailing
list currently numbers more than 30 members. The Wilson and Asqtad
inverters are as fast if not faster than any other software package
available for PC clusters. FermiQCD is an Open Source project and
users can contribute to its improvement by creating new classes and
adding functionality. Some of our objectives include the addition of
an optimized gauge action, optimized Domain Wall fermions, HMC for
dynamical fermions, compatibility with the ILDG format~\cite{ildg},
support for the SciDAC QMP API, and a GUI for visual development.

FermiQCD and additional documentation can be downloaded from:
{\tt www.fermiqcd.net}

\section*{Acknowledgements}

We wish to acknowledge the Fermilab theory group, the University of
Southmpton, and the University of Iowa for their contribution to the
development of FermiQCD.

\end{document}